\documentclass[letterpaper]{article} 

\usepackage{aaai25}
\newcommand{\answerTODO}[1]{\textcolor{red}{#1}} 
\usepackage{xspace}
\usepackage{enumitem}
\usepackage{comment}
\usepackage{tikz}
\newcommand*\circled[1]{\tikz[baseline=(char.base)]{
    \node[font=\footnotesize,align=center,shape=circle,text=white,fill=black,inner sep=0.1pt] (char) {#1};}}

\usepackage{times}  
\usepackage{helvet}  
\usepackage{courier}  
\usepackage[hyphens]{url}  
\usepackage{graphicx} 
\usepackage{natbib}  
\usepackage{caption} 
\usepackage{algorithm}
\usepackage{algorithmic}

\usepackage{booktabs}
\usepackage{subcaption}
\usepackage{xurl}

\usepackage{newfloat}
\usepackage{listings}
\DeclareCaptionStyle{ruled}{labelfont=normalfont,labelsep=colon,strut=off} 
\floatstyle{ruled}
\newfloat{listing}{tb}{lst}{}
\floatname{listing}{Listing}
\title{ Understanding the Complexities of  Responsibly Sharing NSFW Content Online } 
\author {
    Shalini Jangra\textsuperscript{\rm 1},
    Zaid Almahmoud \textsuperscript{\rm 1},
    Suparna De  \textsuperscript{\rm 1},
    Gareth Tyson,\textsuperscript{\rm 2},
    Ehsan-Ul Haq \textsuperscript{\rm 3},
    Nishanth Sastry \textsuperscript{\rm 1}
}
\affiliations {
    \textsuperscript{\rm 1}University of Surrey, United Kingdom\\
    \textsuperscript{\rm 2}Queen Mary University of London, United Kingdom\\
  \textsuperscript{\rm 3}  Hong Kong University of Science and
Technology,
Guangzhou, China\\
    s.jangra@surrey.ac.uk, za0005@surrey.ac.uk, s.de@@surrey.ac.uk, g.tyson@qmul.ac.uk, euhaq@hkust-gz.edu.cn, n.sastry@surrey.ac.uk

}

\usepackage{bibentry}

\newcommand\gareth[1]{\textbf{\textcolor{red}{GT: #1}}	}
\newcommand\ns[1]{\textbf{\textcolor{magenta}{NS: #1}}	}

\newcommand\sd[1]{\textbf{\textcolor{purple}{SD: #1}}	}

\newcommand\Zaid[1]{\textbf{\textcolor{brown}{ZA: #1}}	}

\renewcommand\gareth[1]{}
\renewcommand\ns[1]{}
\renewcommand\sd[1]{}
\renewcommand\Zaid[1]{}

\begin{document}
\maketitle
\begin{abstract}
Reddit is in the minority of mainstream social platforms that permit posting content that may be considered to be at the edge of what is permissible, including so-called Not Safe For Work (NSFW) content. However, NSFW is becoming more common on mainstream platforms, with X now allowing such material.
We examine the top 15 NSFW-restricted subreddits by size to explore the complexities of responsibly sharing adult content, aiming to balance ethical and legal considerations with monetization opportunities.
We find that users often use NSFW subreddits as a \textit{social springboard}, redirecting readers to private or specialized adult social platforms such as Telegram, Kik or OnlyFans for further interactions. They also directly negotiate image ``trades'' through credit cards or payment platforms such as PayPal, Bitcoin or Venmo. Disturbingly, we also find linguistic cues linked to non-consensual content sharing. To help platforms moderate such behavior, we trained a RoBERTa-based classification model that outperforms GPT-4 and traditional classifiers, such as logistic regression and random forest, in identifying non-consensual content sharing, showing better performance in this specific task. The source code and model weights are available at \url{https://github.com/socsys/15NSFWsubreddits}.

\end{abstract}

\noindent\textbf{\textcolor{red}{Trigger Warning:}} \textit{This paper contains references to sensitive topics. Reader discretion is advised.}

\section{Introduction}

Explicit content is prevalent across the Web --- Adult-related websites host \textit{user generated} explicit content~\cite{tyson2016measurements}, and some also incorporate social networking features~\cite{tyson2015social}. 
This is not illegal in many jurisdictions, although it is common to have age restrictions (e.g., 18 years or older). Since such content can, even if legal, be seen as objectionable by many, the acronym NSFW (Not Safe for Work) is often used as a trigger warning for material that is inappropriate for public or professional environments \cite{cauteruccio2022extraction}. 

One reason for the huge amount of adult content is its profitability. The online adult entertainment industry was estimated to have grown larger than Netflix, and all of Hollywood in 2018~\cite{guardian}, and the growth continues in 2024 as well~\cite{yahoo}. 
Regardless of the monetary incentive, mainstream platforms have generally avoided hosting such content. For instance, social networks such as Facebook have policies restricting NSFW content~\cite{fbpolicy}. 

However, new policy changes on X (formerly Twitter) mean that users are able to share consensually produced explicit content that is properly labelled~\cite{Xpolicy}.
While this might open up new monetization possibilities for X, it also brings potential challenges regarding online harms.
Given these potentially huge implications, we  wish to understand  \textit{what might be the role for mainstream platforms in sharing NSFW content, and whether hosting such content sharing can challenge common norms of mainstream platforms around responsible and consensual sharing}.

To this end, we turn to Reddit, which occupies a unique position as being a well-known and widely used platform, yet having a highly flexible and permissive policy around content. Each so-called subreddit is allowed to define its own content policy for its user community, which is then (supposed to be) enforced by its moderators. Also, Reddit does have a global set of rules~\cite{reddit-rules}, which supersede subreddit policies and require that the content shared be legal \cite[Rule 7]{reddit-rules}, consensually posted  \cite[Rule 3]{reddit-rules} and have appropriate labels for explicit or graphic content \cite[Rule 6]{reddit-rules}. However, prior work \cite{matias2019civic} highlights the unique emotional and ethical labor involved in moderating sensitive content, with NSFW subreddits often presenting distinct moderation challenges such as content ambiguity, platform scrutiny, and heightened reputational risk.

\begin{figure}[t]
    \centering
        \includegraphics[width=.95\columnwidth, trim={1.72cm 1.32cm 0cm 0.7cm}, clip ]{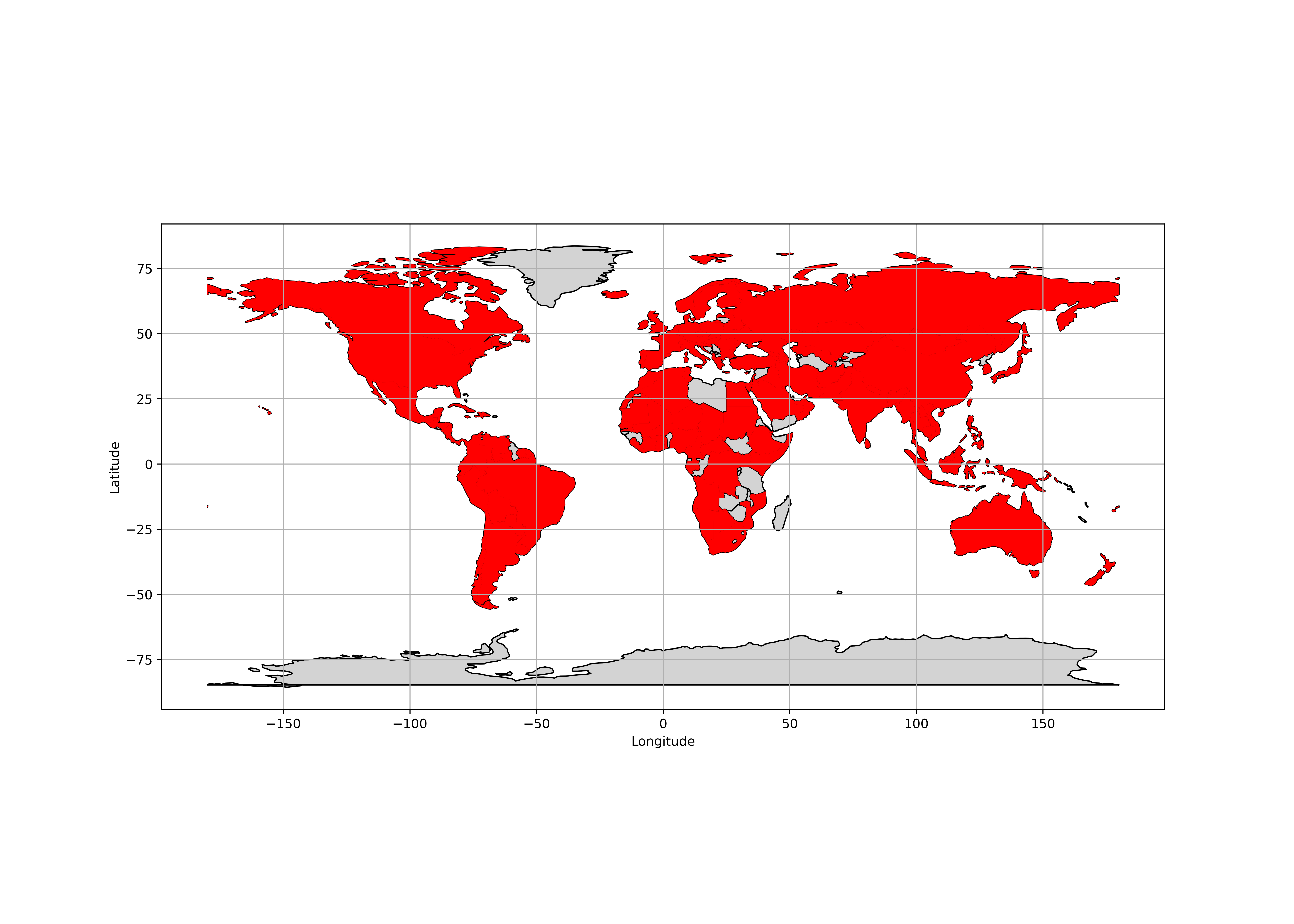}
        \caption{Locations of posters to Top 15 restricted NSFW subreddits studied in this paper. ($\approx$ 155 countries included)}
        \label{fig:map}
\end{figure}

There are several NSFW subreddits that permit explicit content. We focus on so-called \textit{restricted} subreddits where only authorised users are allowed to post. In theory, users allowed to post in restricted subreddits would have been vetted by the moderators, allowing us to analyze the complexities of \textit{responsibly} sharing NSFW content in controlled digital spaces. Analyzing these subreddits provides insight into how a mainstream platform navigates the challenges of managing NSFW content within restrictive environments.

We approach our study by first obtaining ethics approval and collecting  {nine years of data, from 2016--2024. 
To identify appropriate subreddits, we start with 47,184 NSFW subreddits indexed at }\url{https://chaturbot.co/reddit_nsfw_index}, which we accessed on March 18, 2025.
We then select all restricted (this is noted in the ``special'' column) NSFW subreddits with at least 1 million subscribers. This results in 15 communities (Table~\ref{tab:subreddits}) ranging in size from $\approx$ 1 million to over 3 million subscribers from across all six inhabited continents (see Fig.~\ref{fig:map}). 

Exploiting the above data, we answer the following research questions around responsible sharing of NSFW content: 
\begin{enumerate}[label=\textbf{RQ\arabic*}]
    \item \textit{How do NSFW subreddits function as gateways to more specialized NSFW platforms?} 
    This is important to users because they may end up with the additional cognitive burden of having to maintain a profile on ``yet another platform''~\cite{zhong2017wearing}. 
    
    \item \textit{(How) do users make money?} Monetization is an important incentive for many users and also platforms, but may come with additional legal~\cite{tarrant2016pornography} or political~\cite{waltman2021pornography} considerations. 
    
    \item \textit{Are rules around consensual sharing respected?} This has become a pressing question for platforms, regulators and law enforcement, as well as an important research consideration for criminologists and legal scholars~\cite{noncons1,noncons4}, psychologists~\cite{noncons1} and many social scientists~\cite{noncons3,noncons2}.
\end{enumerate}

We identify many interesting practices: Many users use the NSFW subreddits as a \textit{social springboard}, and refer their readers to their profiles on specialized platforms such as OnlyFans~\cite{vallina2023cashing}, which are designed around monetization of adult content, or invite them to connect on private messenger-like services such as Kik and Telegram, where presumably, they can offer more bespoke services or communications, again with a potential monetary angle. Such redirection allows users to focus their attention on developing their presence on one or two platforms whilst benefiting from additional publicity and exposure on NSFW subreddits. However, such behaviour violates the specific rules of many of the subreddits we study, which explicitly prohibit advertising, redirecting users to other chat platforms like Kik or OnlyFans, and engaging in sales activities. The community guidelines emphasize that social media spamming and selling are not allowed, as they undermine the subreddit's primary purpose of content sharing.

We also find posts which \textit{directly trade content}, using credit cards and payment platforms such as PayPal or Bitcoin, often mentioning explicit \$ values. To promote such posts, users often engage in coordinated campaigns, posting the same content from multiple accounts and also sharing the same content across different subreddits. Unfortunately, a number of posts appear to promote malicious or illegal content, which could include illegal file-shares, unauthorized adult content etc.\, which may violate platform guidelines or local laws.

Finally, we uncover disturbing evidence of a small but nonetheless significant number of posts engaged in what might be \textit{non-consensual sharing}. We identify such non-consensual posts first from a manual examination and develop RoBERTa-based classifiers \cite{liu1907roberta} to detect and classify such posts. Our model achieves a Recall score of of 0.86. In alignment with the protocol approved by our ethics review panel, we reported these posts to an organization helping Revenge Pornography victims, and upon their advice, we have also informed local law enforcement. Our model is made available freely to help moderators, platforms, and regulators detect non-consensual sharing practices at  \url{https://github.com/socsys/15NSFWsubreddits}. 

\section{Related Work}
\label{sec:rw}

Early studies on adult multimedia services \cite{tyson2015people,tyson2016measurements,tyson2013demystifying}, often referred to as ``Porn 2.0,” show how traditional adult content platforms have adopted Web 2.0 features, allowing for user interaction through likes, comments, and uploads. Research on platforms like YouPorn demonstrates the extent of user engagement, content variety, and global traffic patterns \cite{tyson2016measurements}. This shift parallels mainstream social platforms, though the demographic and interaction trends exhibit notable differences, particularly regarding gender and user engagement.

Highlighting the need for technology-driven solutions to detect non-consensual content,
work by Mohanty et al. \cite{mohanty2019photo} proposes a deep learning-based system for identifying revenge pornography images, combining nudity detection with face recognition to track down such images across platforms. This is complemented by broader research on non-consensual pornography, such as a systematic mapping study \cite{falduti2023mapping} that categorizes technical approaches to combating this issue and identifies gaps in current responses, particularly within the fields of cybercrime and digital forensics. 

Several studies examine economic activities tied to explicit content, ranging from illicit to institutionalized forms, such as 
the “eWhoring” phenomenon \cite{pastrana2019measuring}, where perpetrators simulate online sexual encounters for financial gain through fraudulent use of sexual images, showcasing how offenders monetize explicit content, and a study of the OnlyFans platform \cite{vallina2023cashing}
that offers a legitimate and highly structured ecosystem where content creators directly engage with consumers to cultivate followings across various social media.

Reddit, one of the few mainstream platforms explicitly permitting NSFW content, has seen a rise in research focusing on its adult communities, including those that emphasize the unique social dynamics and network cohesion within NSFW communities in comparison to SFW posts \cite{corradini2021investigating}, as well as text analysis revealing language cues and community structure in user interactions within NSFW subreddits \cite{cauteruccio2022extraction}.

Beyond content sharing, studies have explored societal and psychological implications of explicit content distribution, including that of AI-driven content.
Felmlee \textit{et al.}'s mixed-methods analysis of gendered aggression on Twitter reveals how sexist language reinforces traditional stereotypes and norms, suggesting that such language is often weaponized to control and demean women \cite{felmlee2020sexist}, while Gamage \textit{et al.} explore public sentiment and ethical concerns raised by the proliferation of AI-generated explicit material by studying Reddit deepfake content \cite{gamage2022deepfakes}.

The review of existing work underscores the evolution of NSFW content sharing from niche adult websites to mainstream social media platforms, revealing complex social, technical, and economic dimensions. However, none of the existing studies on Reddit specifically address critical aspects of how mentions of mainstream social media accounts facilitate the trading of explicit content, or identify linguistic cues and patterns associated with non-consensual sharing. This paper addresses these open challenges by investigating such patterns within Reddit’s NSFW communities, highlighting the implications for mainstream platforms as they navigate similar content-sharing challenges.



\section{Dataset and exploratory analysis}
\subsection{Data crawling and pre-processing}
\label{sec:data}
Using the ArcticShift API \cite{ArcticShift}, we collected posts from 2016 to 2024 from top 15 NSFW subreddits of restricted nature, as ranked by subscriber numbers on \url{https://chaturbot.co/reddit_nsfw_index} (See Table \ref{tab:subreddits}). These are restricted subreddits where only certain users can post, or comment, or both. Many operate an ``approved submitters list''. Each of the considered subreddits has clearly mentioned community rules about not engaging in self-promotion or selling of content and services.
\begin{table}[]
\caption{Considered subreddits with subscribers count}
\label{tab:subreddits}
\begin{tabular}{@{}ll@{}}
\toprule
\textbf{Subreddit Name}                                               & \textbf{Subscriber Numbers}      \\ \midrule
\texttt{r/AsiansGoneWild}                                             & 3,184,839                        \\

\texttt{r/ass}     &            2,530,334                \\
\texttt{r/porninfifteenseconds}                                       & 2,098,035                        \\
\texttt{r/juicyasians}                                                & 1,932,730                        \\
\texttt{r/curvy}                                                               & 1,932,097                        \\
\texttt{r/Hotwife}                                                    & 1,668,119                        \\
\texttt{r/asshole}                                                    & 1,581,495                        \\
\texttt{r/paag}                                                       & 1,406,135                        \\
\texttt{r/NaughtyWives}                                               & 1,347,068                        \\
\texttt{r/rapefantasies}                                              & 1,251,746                        \\
\texttt{r/WomenBendingOver}                                           & 1,241,615                        \\
\texttt{r/Ebony}                                                      & 1,220,411                        \\
\texttt{r/boobbounce}                                                 & 1,205,557                        \\
\texttt{r/DegradingHoles}  & 1,099,652                        \\
\texttt{r/wifesharing}                                                & 1,076,298                        \\ \bottomrule

\end{tabular}
\end{table}

\begin{figure}[t]
    \centering
        \includegraphics[width=.56\textwidth]{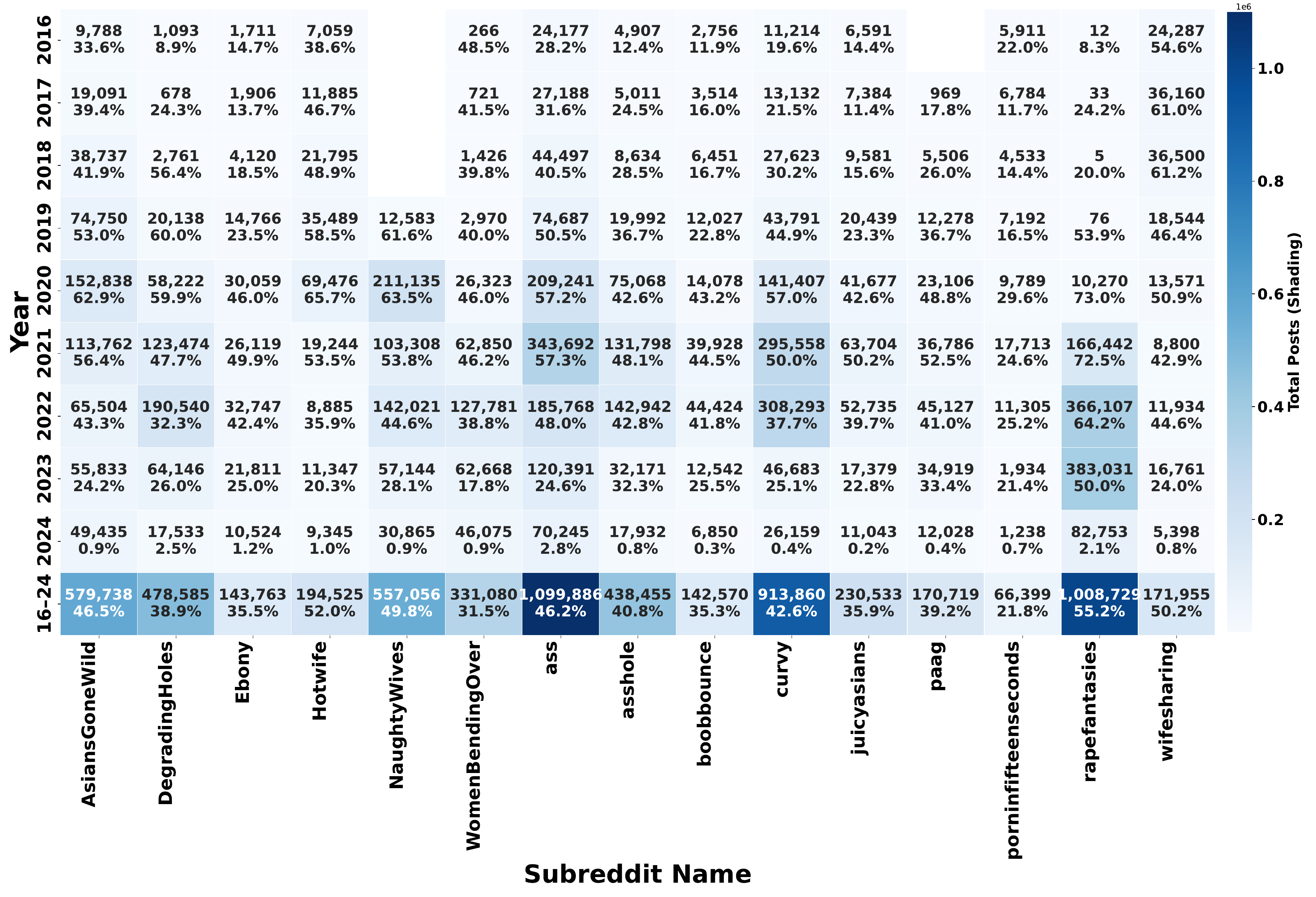}
        \caption{Year-wise distribution of number of posts in subreddits }
        \label{fig:yearwise}
    \end{figure}

\begin{table}[ht]
\caption{Keywords used for content filtering}
\label{tab:keywords}
\resizebox{\columnwidth}{!}{%
\begin{tabular}{lll}
\hline
\begin{tabular}[c]{@{}l@{}}For social\\ media \\ mentions\end{tabular} & \begin{tabular}[c]{@{}l@{}}facebook, fb, instagram, ig, insta, twitter,\\  tweet, tw, snapchat, snap, sc,youtube, yt,\\ tiktok, tt, linkedin, pinterest, pin, whatsapp, \\ wechat, tumblr, telegram, tele, discord, \\ disc, cord, twitch, medium, quora, vine, \\ happn, onlyfans, kik\end{tabular}         & RQ1 \\ \hline
For trading                                                            & \begin{tabular}[c]{@{}l@{}}trade, swap, buy, sell, sale, price, cost, deal,\\  purchase, fee, charge, discount, offer,\\  money, cashapp, pay, apple pay, chime, \\ venmo, dollar, \$, £, bank transfer, PayPal, \\ Zelle, Bitcoin, cryptocurrency, credit card, \\ wire, wallet, revolut\end{tabular} & RQ2 \\ \hline
\begin{tabular}[c]{@{}l@{}}Non\\ consensual\\ sharing\end{tabular}     & \begin{tabular}[c]{@{}l@{}}leak, eleak, rape, voyeur, hidden camera, \\ unaware, not agree, do not know, dentify, \\ porn, hack, hacker, fake,  password\end{tabular}                                                                                                                                        & RQ3 \\ \hline
\end{tabular}
}
\end{table}

To gain a deeper understanding of the data, we first conducted a manual analysis of the randomly selected 5K posts of r/wifesharing subreddit because a lexically similar subreddit (\texttt{r/wifeshar\textbf{e}ing}) has been banned for ``violation of Reddit's rules against non-consensual intimate media and non-consensual sharing". Preparation for manual analysis involved removing offensive content by removing all images and any profane words. We identified profanity by compiling a comprehensive list of profane words from multiple sources: Carnegie Mellon University's `bad words' list \cite{cmu_badwords}, Surge AI's \cite{surge_profanity} profanity dataset, and a collection of profane words from Zac Anger’s GitHub repository \cite{zacanger_profane_words}.

Our analysis uses close reading and exploratory data analysis to identify content-sharing practices within the community. 
This exploratory study reveals three notable patterns: 
\textit{(i)} frequent invitations to connect on social media and private messaging platforms including Kik, Telegram, OnlyFans, and Instagram; \textit{(ii)}  references to content trading and \textit{(iii)} indications of non-consensual sharing. 

Based on our manual analysis, we compiled three categories of keywords (Table~\ref{tab:keywords}): mentions of social platforms (e.g., Kik, FB), keywords indicating trading or money exchange platforms (e.g., PayPal, Bitcoin), and non-consensual sharing (e.g., hidden camera).

This keyword-based approach enabled us to systematically identify and analyze posts in the entire dataset that align with these themes. To identify social media usernames mentioned in those posts, we first filtered posts containing references to social platforms (based on Table~\ref{tab:keywords}) and utilized GPT-4 \cite{peng2023instruction} with few-shot prompting ~\cite{logan2021cutting} (See Appendix \ref{app:prompt}). To evaluate the reliability of our username extraction process using ChatGPT, we conducted a systematic assessment to estimate both false negatives (FN) and false positives (FP). To assess FN, we randomly sampled 100 posts where no usernames were extracted and manually verified that no valid usernames were missed, yielding an estimated FN = 0 for this sample. For FP assessment, we manually reviewed a subset of extracted usernames and identified several common patterns associated with incorrect extractions, including: entries without alphabetic characters, strings shorter than three characters, age/gender indicators (e.g., [50M], [18FTM]), and emoji strings or stylized user metadata. Based on these observations, we implemented targeted post-processing filters to remove such entries. These refinements reduced the number of distinct extracted usernames from 22,515 to 22,108, significantly improving the precision of our dataset. 

Our analysis uncovers the existence of coordinated campaigns within the subreddits, characterized by repeated content sharing and the use of multiple accounts to boost visibility. This suggests  that users may not only be sharing content within the subreddit but also leveraging these campaigns to drive traffic to external platforms for broader engagement or even trading purposes.

Given these patterns, it is crucial to explore whether NSFW subreddits serve as more than just a content-sharing space. In Section~\ref{sec:rq1}, we examine the NSFW subreddit's role as a springboard to other social media platforms, and in Section~\ref{sec:rq2}, we study links to social or other payment platforms that facilitate direct trading. 
\begin{figure*}[t]
    \centering
    \begin{subfigure}[t]{.48\textwidth}
        \centering
        \includegraphics[width=.98\textwidth]{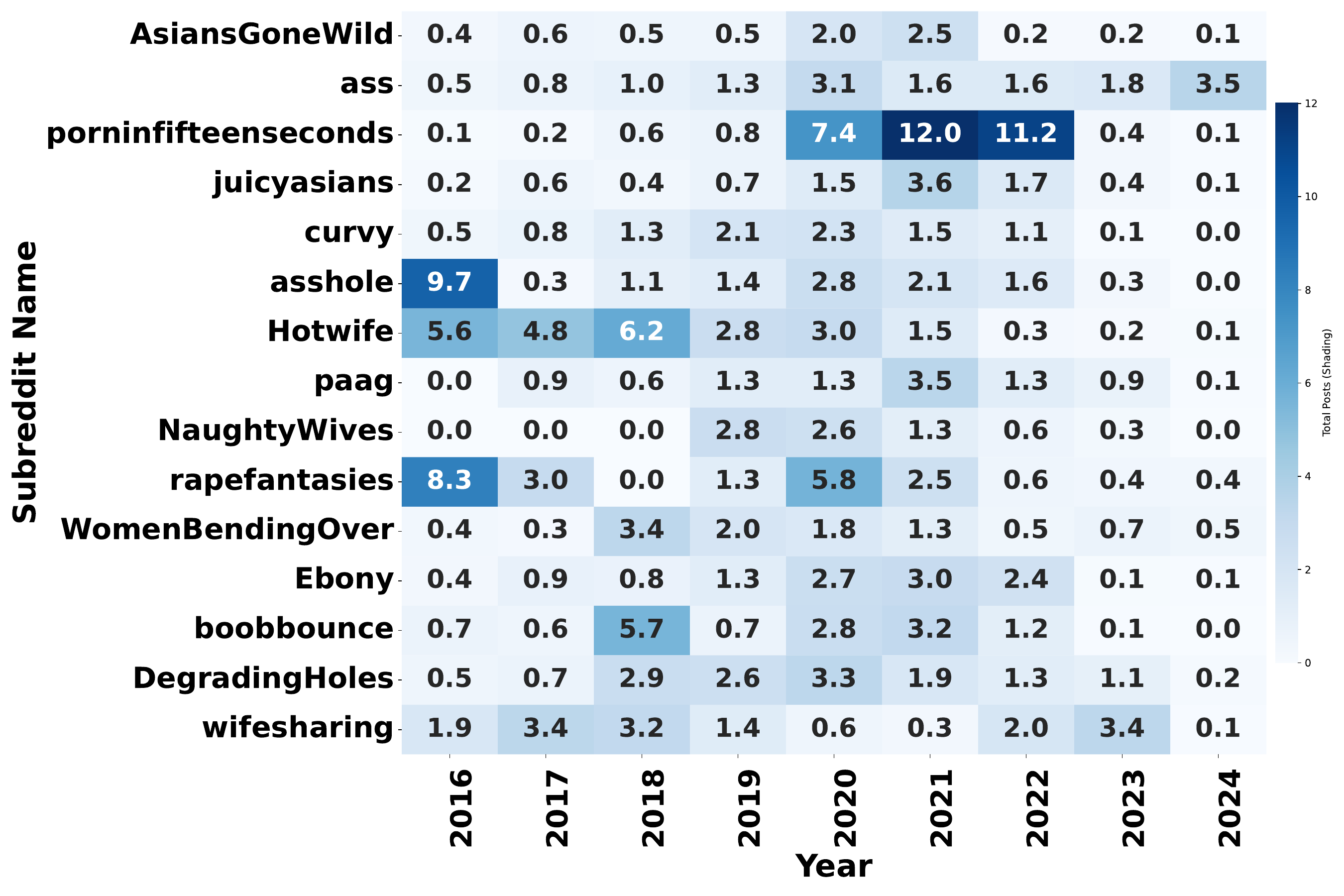}
        \caption{Posts having social media mentions across subreddits.}
        \label{fig:post_percentage_sm}
    \end{subfigure}
    \quad
    \begin{subfigure}[t]{.48\textwidth}
        \centering
        \includegraphics[width=.98\textwidth]{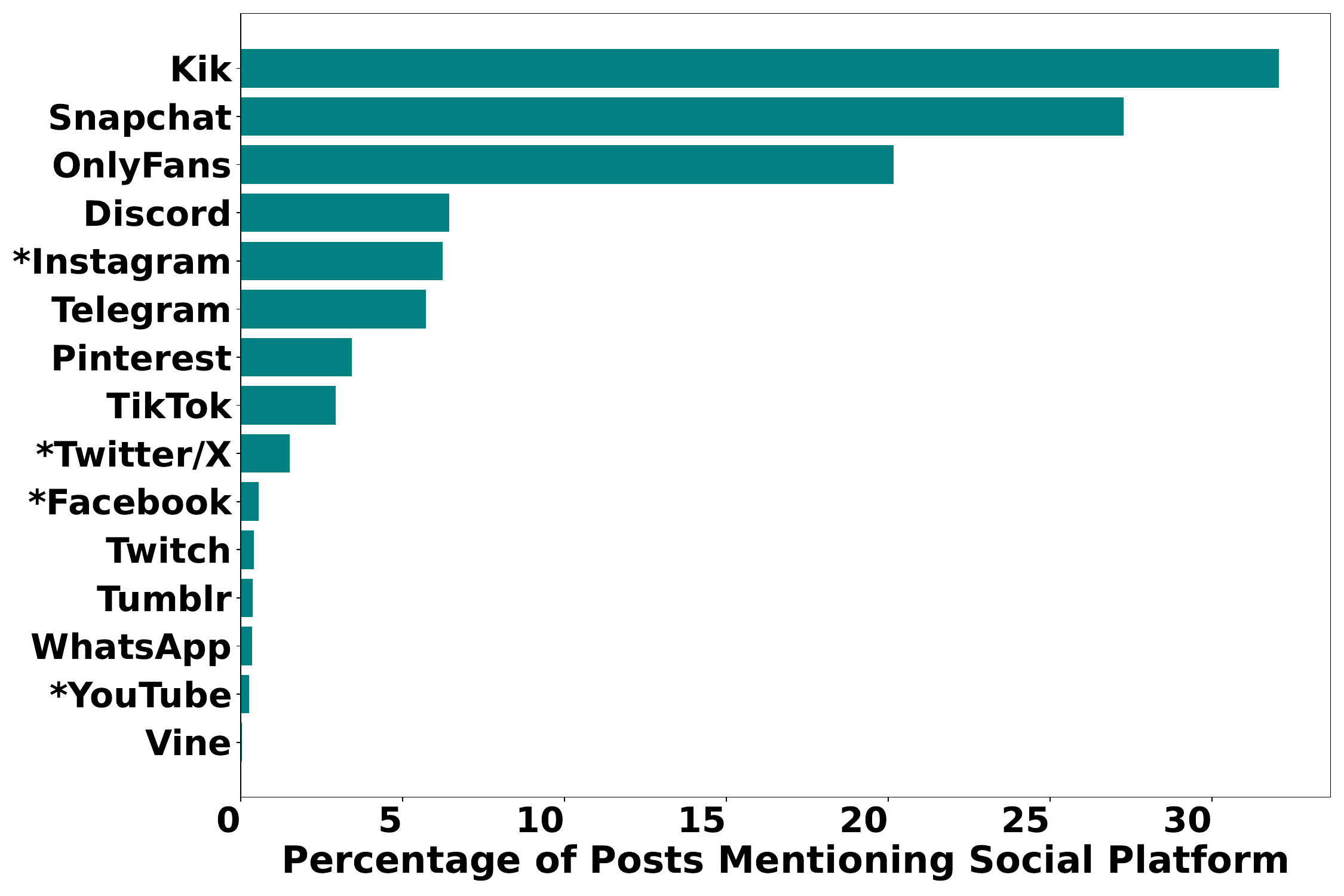}
        \caption{Top social platforms mentioned in the posts }
        \label{fig:platform}
    \end{subfigure}
    \caption{Social media engagement trends in  NSFW subreddits from 2016--24. (a) compares the overall percentage of posts mentioning user handles on social media. (b) reveals the top 15 platforms mentioned. An asterisk indicates public platforms like Instagram and Twitter, often used for public feeds, in contrast to private platforms like WhatsApp or Telegram.}
    \label{fig:combined_social_media_analysis}
\end{figure*}

\subsection{Annotation of non-consensual posts}
\label{sec:annotate}
Non-consensual sharing of intimate content is a pressing issue, especially on social media platforms where privacy may be breached. Detecting posts that may involve such sharing requires a deep understanding of the language and context in which this content is conveyed. 

Given the extensive size of the dataset, manual annotation proved to be labor-intensive. 
So we selected 5 random subreddits out of the 15 in our dataset for manual annotation: \texttt{r/curvy}, \texttt{r/Hotwife}, \texttt{r/Naughtywives}, \texttt{WomenBendingOver} and \texttt{r/wifesharing}.
To streamline the process, we derive keywords (see Table~\ref{tab:keywords}) indicating non-consensual posts from manual analysis. Moreover, we observe that users often use intentional misspellings, such as "le@k" for "leaks" and "R4PE" for "rape," to evade detection. These modifications are part of leet speak (1337 speak) \cite{leet_wikipedia} that typically involves replacing letters with visually similar numbers or symbols to bypass content moderation. To address this challenge, we normalize the text by converting such symbolic variations back into their standard forms before applying keyword filtering.
This step identified 3718 posts, which were then manually annotated by three annotators. Three independent annotators evaluate the posts, labeling them as ${0:consensual}$ or ${1:non-consensual}$. 298 posts are labeled as non-consensual by at least two annotators. We later use this dataset to train a RoBERTa-based model to detect non-consensual sharing  (Section~\ref{sec:classifier_new}). 
\begin{figure*}[!t]
    \centering
\begin{subfigure}[t]{.45\textwidth}
        \centering
        \includegraphics[width=.98\textwidth]{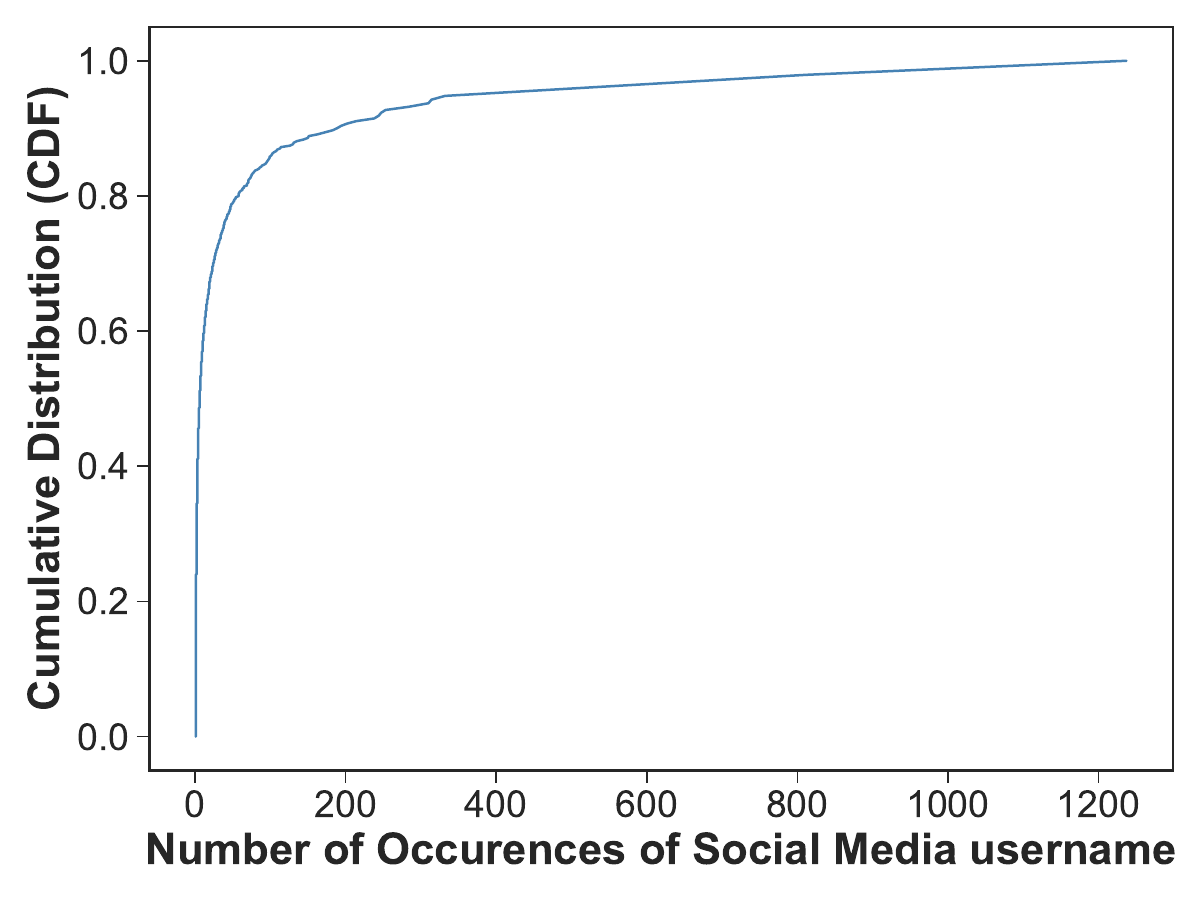}
        \caption{Cumulative Distribution Function (CDF) of username mentions per subreddit.}
        \label{fig:usernamecdf}
    \end{subfigure}
     \quad
\begin{subfigure}[t]{.43\textwidth}
        \centering
        \includegraphics[height=6.3cm]{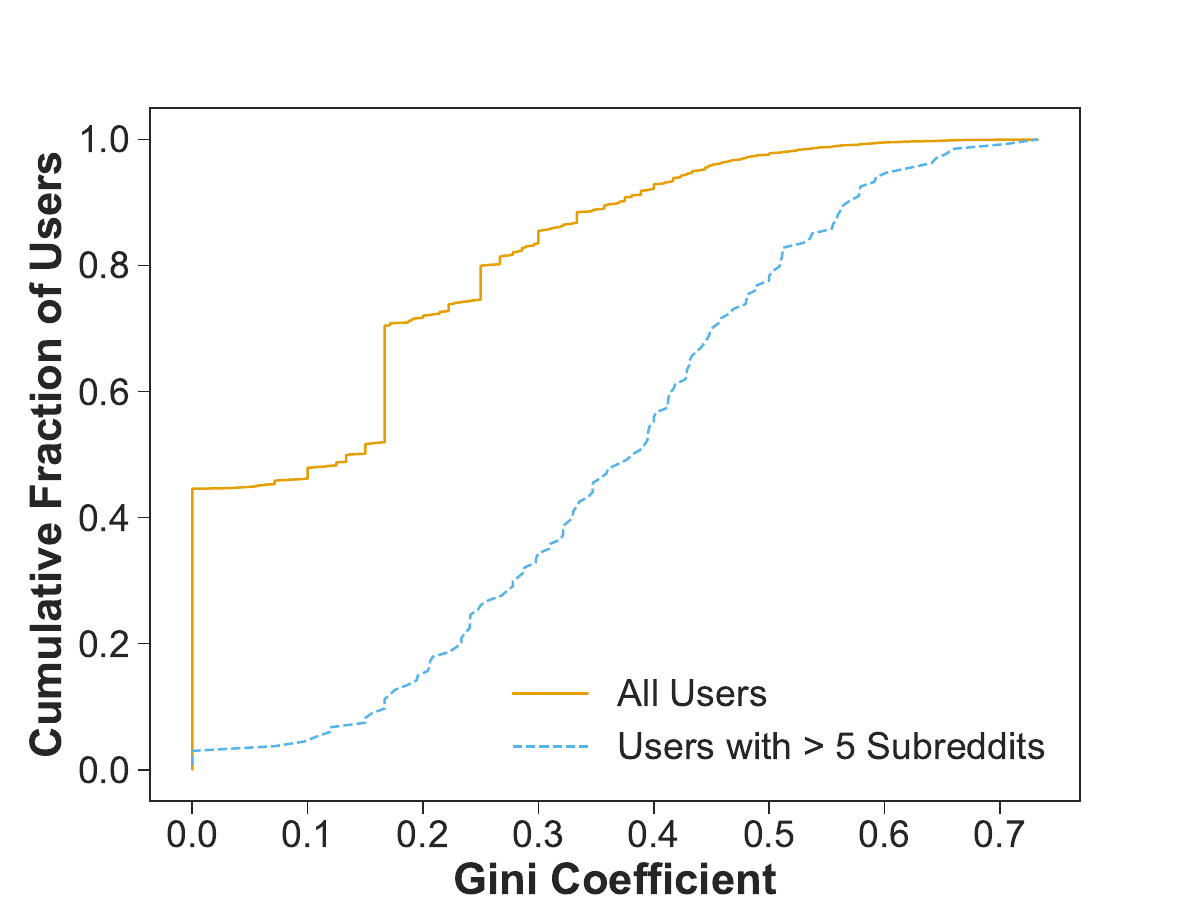}
        \caption{CDF of Gini Coefficients Across Users.}
        \label{fig:lorentz}
    \end{subfigure}
 \quad 
    \begin{subfigure}[t]{.45\textwidth}
        \centering
        \includegraphics[width=.98\textwidth]{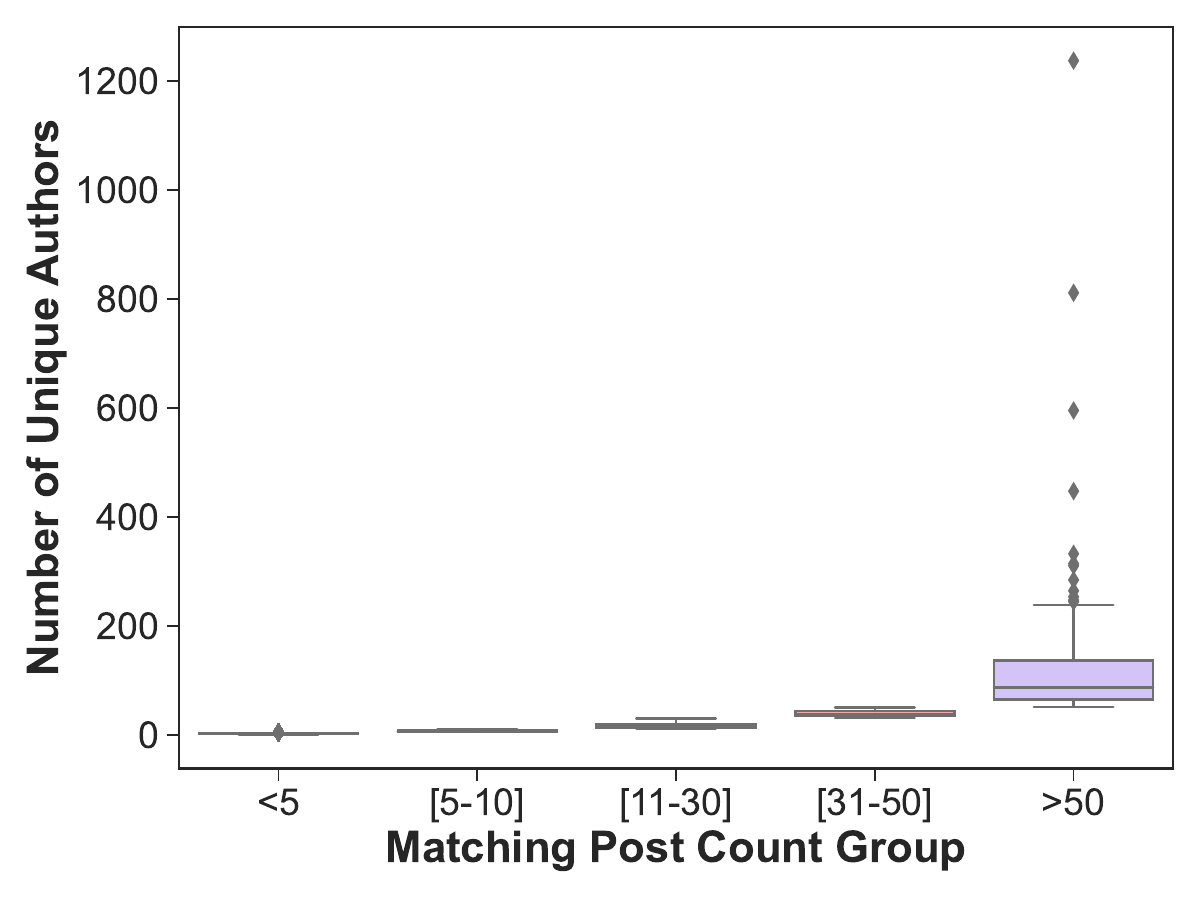}
        \caption{Author distribution per campaign.}
        \label{fig:campaign_diff_authors}
    \end{subfigure}
    \quad
    \begin{subfigure}[t]{.45\textwidth}
        \centering
        \includegraphics[width=.98\textwidth]{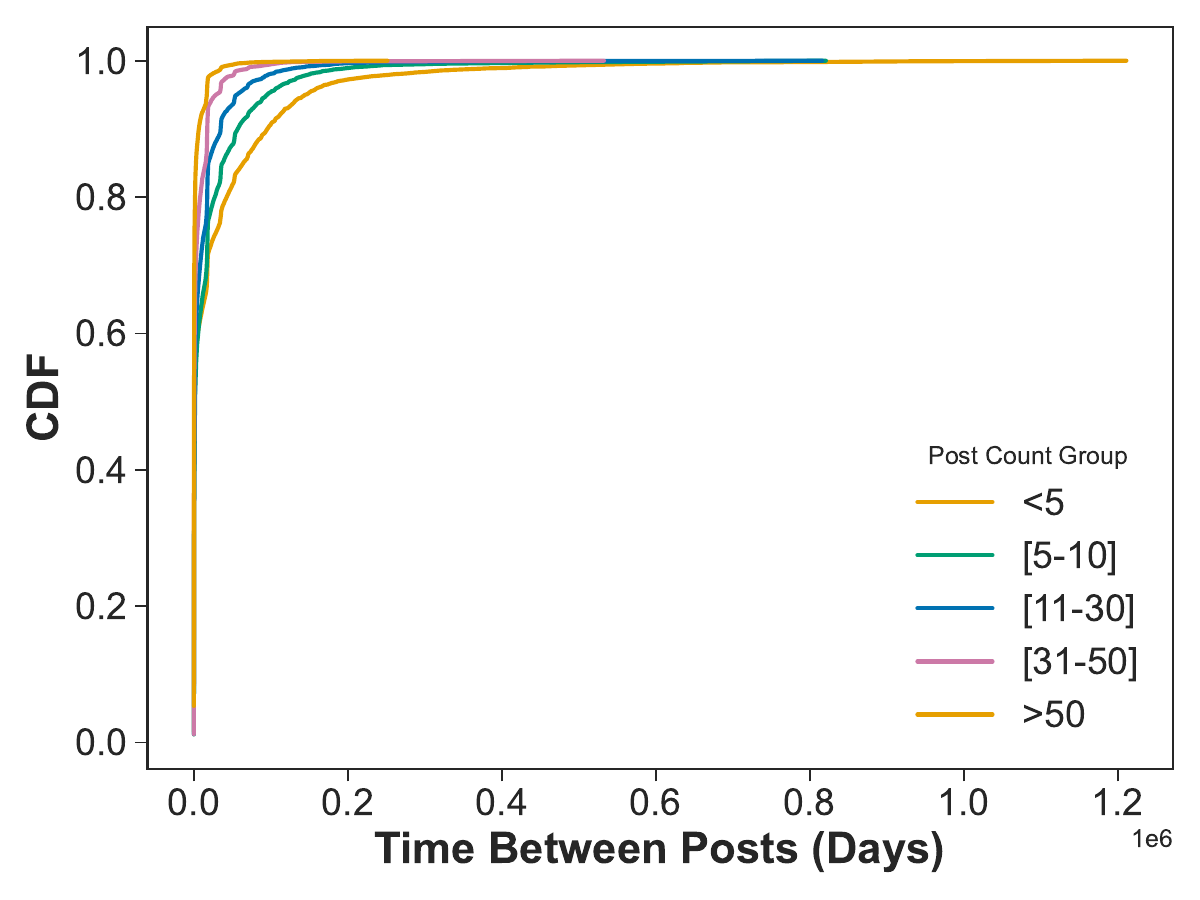}
        \caption{Campaign duration}
        \label{fig:campaign_duration}
    \end{subfigure}  
    \caption{Illustrate the distribution, authorship, and temporal dynamics of social media username mentions in Reddit posts. (a) shows that while most usernames are rarely repeated, a small fraction dominates the space. (b) reveals a similar inequality in subreddit participation, with a small user base contributing the majority of activity. (c) suggests that high-frequency usernames are often posted by many different users, hinting at viral or coordinated behavior. (d) supports this further by showing compressed posting time windows for these high-frequency usernames—consistent with campaign-driven or automated posting patterns.}
    \label{fig:campaign}
\end{figure*}
\section{RQ1: NSFW subreddits as Social Springboards}
\label{sec:rq1}

Despite rules against ``social media spamming,” a notable portion of all subreddit's posts contains references to external social media platforms, which reflects the subreddit's role as a gateway to other communication channels. Rather than serving as an endpoint, NSFW subreddits therefore appear to function as a springboard, directing users to other platforms where interactions may be more private, personalized, or unmoderated. This phenomenon underscores the subreddit's unique role within the digital ecosystem as an initial point of contact that facilitates extended interactions on other platforms.

\subsection{Mentions of non-Reddit social handles}

Fig.~\ref{fig:post_percentage_sm} displays NSFW subreddit activity over time (2016–2024), showing the relative proportion (\%) of posts per subreddit per year mentioning social media usernames from platforms other than Reddit. This started early in some subreddits such as \texttt{r/asshole}, \texttt{r/rapefantasies}, and \texttt{r/hotwife}, which showed sharp spikes in 2016 (e.g., 9.7\% , 8.3\% and 5.6\% respectively). However, it is important to note that the total number of posts in these subreddits was relatively low in 2016 (Refer Fig \ref{fig:yearwise}), so even a small increase in activity could result in a disproportionately high percentage. This trend was not maintained in subsequent years, likely due to the expansion of the NSFW ecosystem on Reddit, where a growing number of subreddits diluted the relative share of social media mention activity in these early subreddits. We observed an overall increase in social media mentions during 2020 and 2021 across all subreddits. We believe this can be attributed to a massive increase in online activity and adult content consumption during the COVID-19 lockdown \cite{zattoni2021impact}. The observed peak of social handle mentions in \texttt{r/pornin15seconds} (12\% in 2021) is due to the increased popularity of that subreddit due to viral or trending format videos (quick clips) that gained attraction during this time, aligning with TikTok-style short-form content growth that influenced cross-platform attention. 
This pattern of usage of mentions of handles from other social platforms  suggests an evolving role for Reddit as a \textit{social springboard} facilitating social media interaction by directing users toward alternative platforms. 
\subsubsection{Social media username mentions}
Overall, 68.5\% of posts mentioning social media platform names include user handles, indicating a more direct sharing of contact information. For posts that do not explicitly share usernames, users often employ indirect phrases to convey username information, such as ``ping me for username" ``tele (same name)" or instructions like ``DM kik/tele in bio" and ``Kik same as username".
This suggests that even without directly stating usernames, users find creative ways to communicate details, facilitating interactions off-platform. This behaviour highlights a growing tendency among users to work around community guidelines to promote their adult social media profiles and facilitate off-Reddit interactions.

Fig.~\ref{fig:platform} illustrates the top 15 social platforms mentioned. An asterisk preceding a platform name denotes a public platform, which is typically used for more visible, wide-reaching interactions. Public platforms like Instagram, X, and Facebook primarily serve to showcase public personas. Private platforms, such as WhatsApp, Telegram, Snapchat, and OnlyFans, cater to different user intents by providing more controlled and secure environments for one-on-one or small-group interactions, which can potentially be monetized. Kik emerges as the dominant platform, accounting for  over 30\% of mentions,  
likely due to its ease of use, emphasis on anonymity, and its suitability for less moderated content.

\subsection{Campaign-like patterns in username sharing}
Fig.~\ref{fig:usernamecdf} shows the cumulative distribution function (CDF) of social media username mentions across the subreddits we study. Each point in the distribution is the number of times a distinct social media username was mentioned within a single subreddit. The steep rise in the CDF at lower occurrence values indicates that most usernames are mentioned only a few times, but a small fraction of usernames are mentioned significantly more frequently. 
Thus, the promotion of other social media handles in any given subreddit is dominated by a few active players. 

However, we also found that many users are \textit{posting the same username  across different subreddits}. Indeed, social handles are mentioned on 14 out of the 15 subreddits we consider. Thus, although the number of mentions of a social handle may be few within a single subreddit, it could also be part of a larger \textit{campaign} crossing multiple subreddits.

To see whether such campaigns are focusing equally on all the subreddits, for each social username mentioned, we compute a Gini coefficient. Gini coefficients~\cite{catalano2009gini} are used to quantify how unequally a resource is distributed, with a value close to 1 indicating an equal distribution and a value closer to 0 indicating highly unequal distribution. Here, we are interested in seeing how unequally a username's mentions are scattered among the different subreddits it is mentioned in. Fig.~\ref{fig:lorentz} shows that when considering all users, most of them show highly unequal distribution (i.e., Gini close to 0, indicating that they focus mostly on a handful of subreddits), whereas for users who share their usernames in more than 5 subreddits, they mention their username in more equal numbers across the different subreddits they participate in, indicative of a more coordinated campaign of sharing their social media usernames for social springboarding to other platforms.

\subsubsection{Reddit Authors/Accounts Distribution per Campaign }
Figure \ref{fig:campaign_diff_authors} displays a box plot showing the distribution of the number of unique Reddit authors/accounts associated with social media usernames mentions, grouped by the frequency of those mentions across posts.  A key observation is the stark contrast in the distribution for the $>$50 post group. While most username mentions in the lower-frequency groups ($<$5, [5–10], [11–30], [31–50]) are typically associated with very few unique authors—indicating single-user promotion or occasional mentions—the $>$50 group shows a substantial increase in the number of distinct authors. Some of these usernames are mentioned by hundreds or even over a thousand different Reddit accounts. This suggests a likely pattern of coordinated or viral activity. In particular, usernames or account identifiers in the $>$50 group may represent widely shared promotional accounts, such as influencers, OnlyFans creators, or content services and mass marketing campaigns, where multiple users or bots advertise the same account.
\subsubsection{Campaign Duration and Group Size}
Figure \ref{fig:campaign_duration} presents a Cumulative Distribution Function (CDF) of the time gaps (in days) between posts for having the same social media usernames.
Tighter Curves for Higher Post Counts, especially [31–50] and $>$50, exhibit significantly steeper CDFs. This indicates that most of their posts occur within short time intervals.
 The purple curve representing usernames mentioned in more than 50 posts rises rapidly, indicating that the vast majority of these posts are made within a narrow time window. This is indicative of coordinated posting behavior, indicating automated or campaign-driven posts.
 In contrast, usernames mentioned fewer than 5 times ($<$5 group) show a broader distribution. Their curve ascends more gradually, suggesting that these mentions are sporadic and distributed over longer time periods—more typical of organic or casual user behavior.
The high concentration of posts in short time frames for frequent usernames points to promotional bursts and cross-account coordination, with different Reddit users posting similar content around the same social media username or handle in a synchronized fashion.

\noindent\textbf{Implications}: Overall, our analysis reveals that NSFW subreddits act as more than just discussion forums. Instead, they serve as gateways to other social media platforms, extending interactions beyond Reddit, with evidence of campaign-like patterns to drive such out-of-platform engagement. This positions the NSFW subreddits as key nodes within a larger network of social media engagement, where users can pursue connections aligned with their preferences for privacy, anonymity, or public exposure. However, this dynamic may also facilitate interactions with potential implications for monetization, fostering a marketplace for transactional exchanges that could bypass platform regulations. Furthermore, by guiding users toward platforms with limited oversight, these subreddits inadvertently open pathways for behaviours that may lead to illegal activities. This evolving role as a connector to both public and private platforms highlights the potential to influence user behavior in ways that raise ethical and legal concerns within the digital landscape.


\section{RQ2: Trading on NSFW subreddits}
\label{sec:rq2}

Turning to the question of whether users use NSFW subreddits to monetize their content, we find substantial evidence indicating trading.
This involves exchanging media, often for money or some reciprocal reward.
While this trend may seem ``obvious'', especially with the previous mentions of promotions for Kik, Telegram or Only Fans, it is important to contextualize our findings within the broader dynamics of the community. In particular, this section looks at monetization directly on Reddit.

\subsection{Overview of Trading}

\subsubsection{Identifying trading}
We first use the keywords mentioned in Table~\ref{tab:keywords} to extract posts related to trading. Recall that this list was formulated based on prior observations of the common language used within the community.

\subsubsection{Frequency of trading}
We identify that 1.12\% of all collected posts across the 15 targeted subreddits contain references to trading activity. While this percentage may initially appear low, it corresponds to a substantial absolute count of 72,868 posts. Of these, 48,292 posts are non-deleted, while 24,576 posts are deleted. It is important to note that for deleted posts, only the titles are available; the full post bodies are not retrievable. Consequently, the actual prevalence of trading-related content is likely underestimated, as additional mentions may have appeared in the bodies of deleted posts. This limitation suggests that the observed 1.12\% figure could represent a conservative lower bound on trading-related activity within these subreddits as quite significant percentage of posts are deleted over the years, refer Fig. \ref{fig:yearwise}.
30.8\% of posts offering the trading via other social media platforms. 
\subsection{Prices}
\begin{figure}
    \centering
    \includegraphics[width=0.49\textwidth]{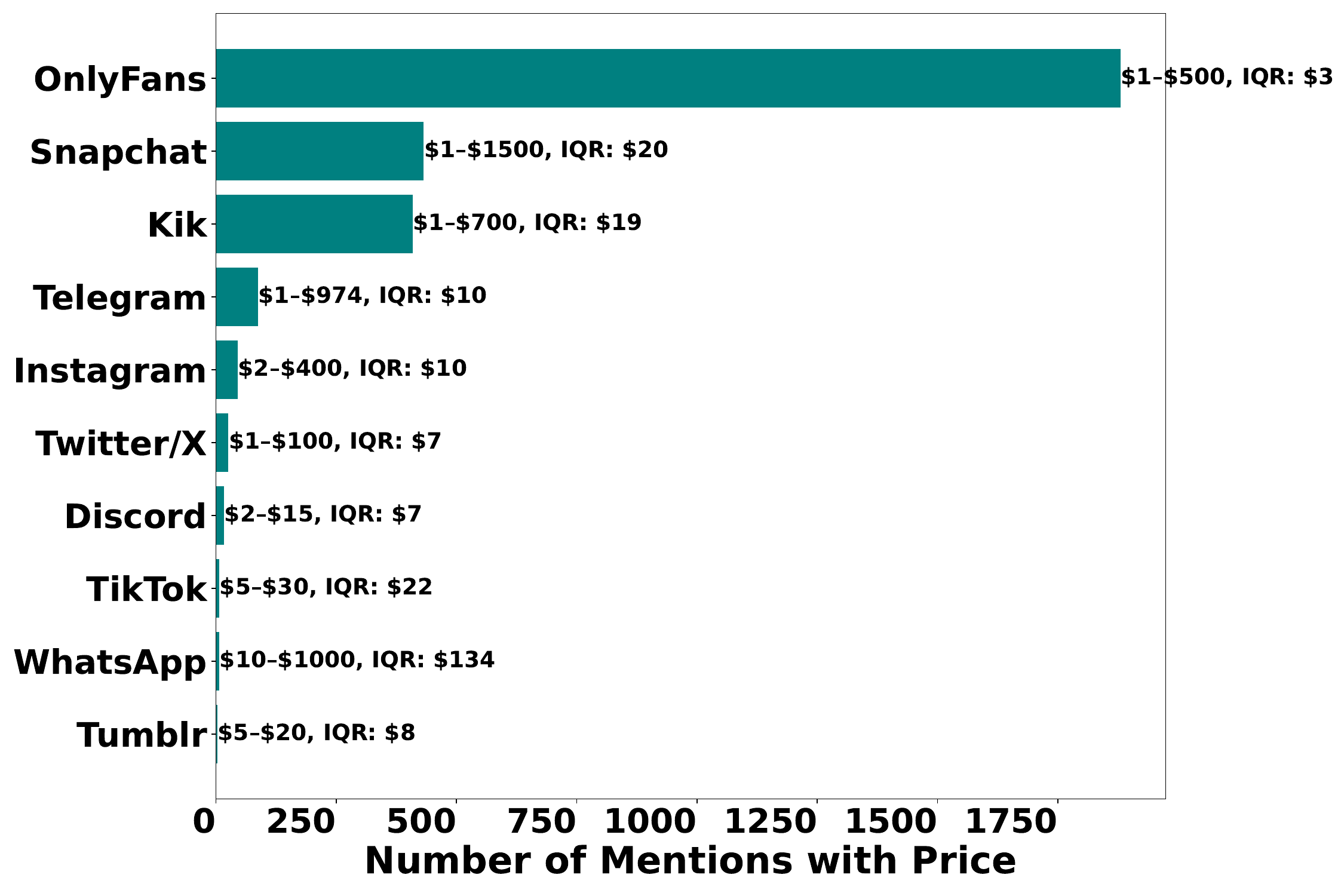}
    \caption{Top 10 Social Media Platforms with Price Mentions (Annotated with Price Range and Inter Quartile Range (IQR))}
    \label{fig:platform_prices}
\end{figure}

We identify 7,430 posts that mention the price of content, with prices going up to US\$1,500. 
Figure \ref{fig:platform_prices} presents the top 10 social media platforms mentioned in posts that include price references. OnlyFans dominates with the highest number of price-linked mentions, followed by Snapchat and Kik. Notably, the price ranges and interquartile ranges (IQR) vary widely across platforms, indicating differing monetization strategies or market norms—for instance, WhatsApp shows high pricing variability (IQR: \$134), while OnlyFans exhibits a narrower price spread (IQR: \$3).
In the remaining price-mentioning posts, users typically either request to be contacted via direct message, share their usernames for other social media platforms, or engage in direct negotiation within the Reddit post itself—for example, posts like "Would you pay \$80 to see a face?" illustrate explicit pricing discussions. Table \ref{tab:payment_apps} presents various payment platforms mentioned across subreddits, along with the number of posts in which each platform is referenced.
\begin{table}[!b]
\caption{payment platforms mentioned}
\centering
\label{tab:payment_apps}
\begin{tabular}{@{}lr@{}}
\toprule
\textbf{Payment App} & \textbf{Frequency of mentions} \\ \midrule
PayPal               & 926                         \\
CashApp              & 903                         \\
Venmo                & 295                         \\
Zelle                & 187                         \\
Revolut              & 35                          \\
ApplePay             & 9                           \\ \bottomrule
\end{tabular}
\end{table}

Table~\ref{tab:Trading_examples} presents examples of posts referencing content trading. 
Post such as “Selling Teen Nudes (16–18),” which suggest the potential circulation of sexually explicit material involving minors on NSFW subreddits. Although the age of consent varies across jurisdictions, Reddit’s policies explicitly prohibit any sexually explicit content involving individuals under the age of 18. Therefore, we reported about the potential presence of sensitive/illegal content to a relevant
charity as well as law enforcement authorities.

\subsection{Trading dynamics}
\subsubsection{ Trading Modes}

\begin{itemize}
    \item \textbf{Gift Card Transactions for Adult Content:} There are mentions of offering adult content access in exchange for Amazon Gift Cards. Example: ``Selling Erito Porn Login + More Want Amazon Gift Cards.''
    \item \textbf{Trading/Negotiation through personal messages:} There are mentions of asking for direct messaging for personalized adult content. Example: ``A mixed album of [m]e and my asian wi[f]e. Do you like our pictures? Feel free to text us, girls are also very welcome :)''
     \item \textbf{Cross-Platform Promotions for Content Sales:} -   Some posts mention multiple social media platforms to advertise and sell adult content, directing potential buyers to preferred payment methods. Example: ``Come buy 5 photos from me for \$12 rn on Twitter!! @ f*i*ds*eci*ls @ j*nn*b*eh on cashapp!!''
     \item {\textbf{Location-Based Adult Services:}  Offering in-person adult services tailored to specific geographic locations, targeting local clientele with personalized offerings. Example: ``In Las Vegas (f) 24 \$600 1 night''.}
\end{itemize}
\subsubsection{ Promotional Strategies}
Sellers employ such strategies to leverage urgency, exclusivity, and affordability to drive subscription growth and maximize engagement. 
\begin{itemize}
    \item \textbf{Time-sensitive Offers:}
A recurring tactic is the use of time-sensitive offers such as discounted subscription rates (e.g., ``\$3/month'' or ``\$4.50 for 30 days''), which appeal to cost-conscious users and stimulate impulse purchases. Scarcity marketing is also evident, with creators offering ``exclusive gifts'' or personalized content to a limited number of early subscribers (``first ten followers''), thereby creating a sense of competition and reward. Additionally, creators frequently emphasize personal interaction and daily activity, framing their pages as intimate and responsive spaces (``very active with fan interactions,'' ``requests taken just for you''), which enhances perceived value.
\item \textbf{Niche Identity Marketing:} Creators often highlight distinctive personal attributes (e.g., ``just started \$8 OnlyFans Argentinian-Asian petite girl'') to attract niche audiences by appealing to specific ethnic, physical, or novelty-based interests within targeted consumer segments.
\item \textbf{Tiered Pricing:} involves offering different price points, ranging from lower-cost options like picture sets for \$5, mid-range video bundles for \$10–\$15, and premium services such as custom content and live interactions for \$20–\$25. 
The creator may employ a value-based, differentiated pricing model in which content is categorized and priced according to its explicitness and focus. For instance, individual images featuring specific body parts, such as ``booty pics,'' are offered at lower price points (e.g., \$5), while more comprehensive and explicit bundles featuring multiple body parts such as breasts, genitals, and bu*t*cks, are priced higher (e.g., \$15). 
\end{itemize}
\subsubsection{Implications:} These observations prompt critical consideration about the evolving nature of interactions within the subreddits. The trading of explicit content raises questions regarding user intentions and the implications for community dynamics, especially as some users leverage their geographic locations to establish credibility or to facilitate transactions in physical rather than cyber space. Furthermore, the actual exchange of money appears to happen off the platform (e.g., via credit cards, PayPal etc.), or on other platforms such as Only Fans or Kik (Section~\ref{sec:rq1}), which impacts the monetising ability of  platforms hosting the NSFW posts.

\begin{table}[]
\caption{Examples of posts referencing content trading}
\label{tab:Trading_examples}
\resizebox{\columnwidth}{!}{%
\begin{tabular}{@{}ll@{}}
\toprule
\textbf{Trading}                                                           & \textbf{Example Post}                                                                                                                                                                    \\ \midrule
Price mentions                                                             & \begin{tabular}[c]{@{}l@{}}Selling Teen Nudes (16-18). No Trading. \\ Dm Prices $10-$100\end{tabular}                                                                                    \\
\begin{tabular}[c]{@{}l@{}}Price and social \\ media mentions\end{tabular} & \begin{tabular}[c]{@{}l@{}}SELLING T33N and NL megas \$8. Dm \\ me here or telegram : \{fe*y*a*ks\}\end{tabular}                                                                         \\
\begin{tabular}[c]{@{}l@{}}Payment app \\ mentions\end{tabular}            & \begin{tabular}[c]{@{}l@{}}YNG/PYT/TEEN/ NL MEGAS FOR \\ SALE ONLY \$7- (PayPal/Cashapp/ \\ Venmo) - for fast reply and instant \\ delivery DM HERE OR TELEGRAM:\\ @p**kt*n\end{tabular} \\
Location mentions                                                          & \begin{tabular}[c]{@{}l@{}}Anyone in or around Maryland that want \\ to *** with us!  Couples/H**g Males/\\ S**y Women? Our inbox is waiting\end{tabular}                                \\
Partner Swapping                                                           & \begin{tabular}[c]{@{}l@{}}Looking to swap pic of my wife for pic \\ of your wife privately\end{tabular}                                                                                 \\
Trading for upvote                                                         & \begin{tabular}[c]{@{}l@{}}19 {[}F4M{]}Looking for fun (1 free nude \\ for upvote)!! snap: he*therdu*n19\end{tabular}                                                                    \\ \bottomrule
\end{tabular}
}
\end{table}

\subsection{Role of other image sharing platforms}
We observe that many of the trading posts also provide pictures, e.g., as samples. Note that the practice is wider than just the trading-related posts, as genuine posts aligned with the original sharing intent of these subreddits also tend to provide images. Interestingly,  with 50.8\% of posts being cross-posted from 9,780 external domains, the remaining 49.2\% have a Reddit-related domain  such as (\texttt{i.redd.it}, \texttt{i.reddituploads.com}).

\begin{figure}[hbt]
    \centering
        \includegraphics[width=.45\textwidth]{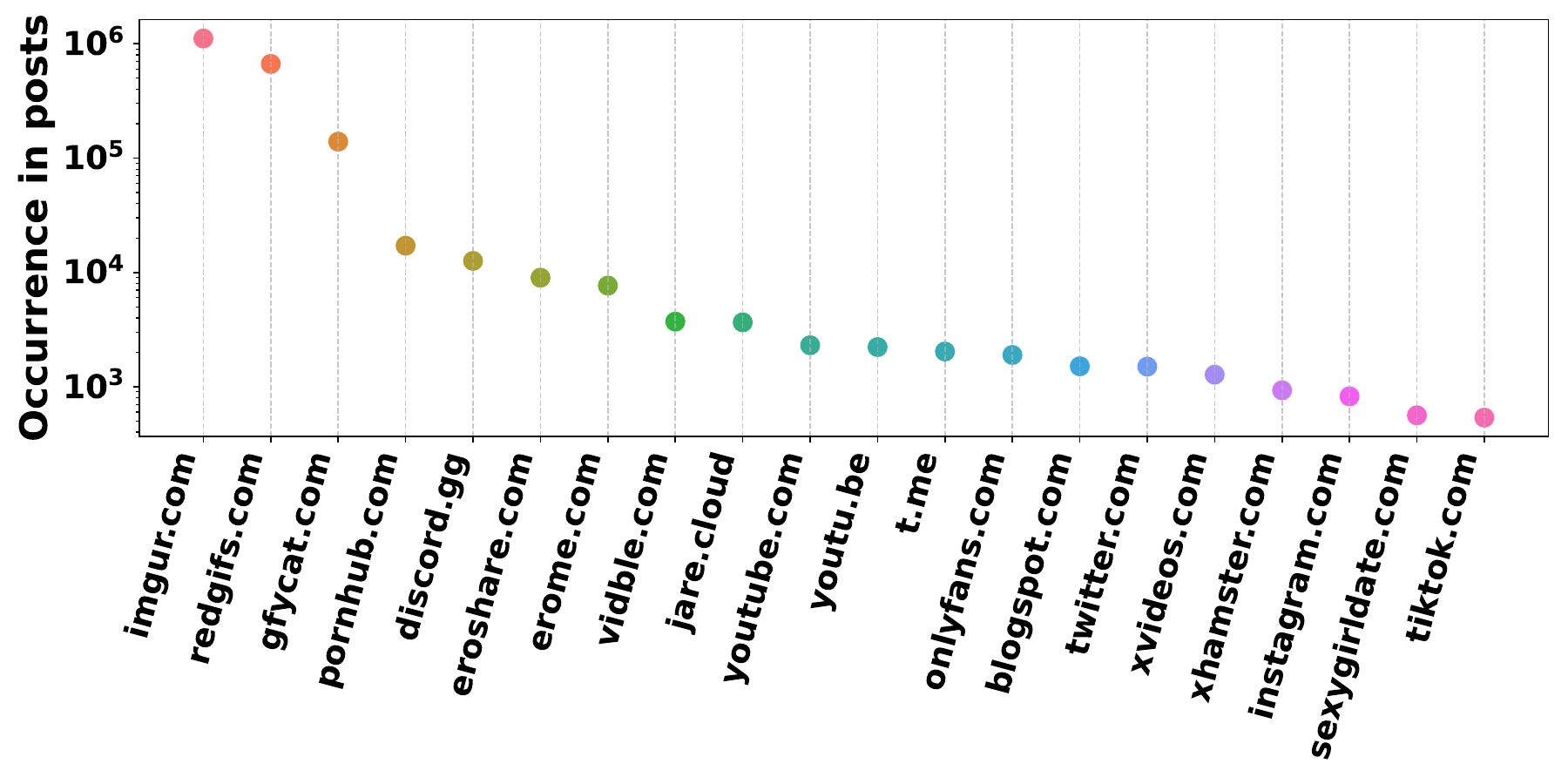}
        \caption{Top 20 domains mentioned across subreddits} 
        \label{domain_frequency}
    \end{figure}

The dot plot in Fig.~\ref{domain_frequency} presents the most frequently referenced external domains (i.e., excluding Reddit’s own infrastructure) across the analyzed subreddits. This distribution highlights the prominence of cross-platform content sharing, where users frequently link to media or profiles hosted on other platforms. 
We combined all the subdomains to root domain such as \texttt{m.imgur.com} and  \texttt{i.imgur.com}) are combined in \texttt{imgur.com}.
Unsurprisingly, the most common domains include major image-hosting sites such as \texttt{imgur.com},  \texttt{redgifs.com}, \texttt{gfycat.com}, and various video-hosting sites, suggesting that visual content is not only central to the community's interactions but also that multimedia sharing forms a significant portion of the subreddit’s activity. Domains such as \texttt{xhamster.com}, \texttt{pornhub.com}, \texttt{onlyfans.com} and \texttt{xvideos.com} also appear on the list, hinting that users  link to or promote content from traditional adult entertainment sites as well.  Social media and identity-extension services like \texttt{twitter.com}, \texttt{instagram.com}, and \texttt{tiktok.com} suggest that users frequently link their broader online presence back to Reddit. In parallel, platforms such as \texttt{t.me} (Telegram) and \texttt{discord.gg} point to the use of private messaging or community coordination tools, often associated with content sharing or solicitation outside Reddit's public threads. The inclusion of \texttt{blogspot.com} and \texttt{youtube.com} illustrates the occasional role of long-form or mainstream media content. This distribution highlights not only Reddit’s media-sharing dynamics but also the platform’s role as a nexus for identity projection, promotion, and off-platform engagement.

\subsubsection{Presence of Link Shorteners}
We found that several domains in the dataset are known URL shorteners or redirection services. 
Some, like discord.gg, youtu.be, and t.me, are commonly used to share links to other platforms, such as Discord servers, YouTube videos, or Telegram channels. Others, like goo.gl, cutt.ly, and forms.gle, are general-purpose shorteners that can be used to link to any type of content. These short links can make it harder to see where the link actually goes, which may be used to hide promotional, adult, or spam content. While some shorteners are used for convenience, others may be used to avoid detection by moderators or content filters. The high number of mentions for certain shorteners suggests they are an important part of how users share and promote content across platforms.
We also noted links to popular online tools like Linktr.ee, allmylinks.com, many.link, etc., which enable users to create a single, customisable landing page with multiple links, and are commonly used to consolidate paid content links (e.g., to OnlyFans, Fansly, or ManyVids).

\section{RQ3: Traces of non-consensual sharing}
\label{sec:rq3}
The rules of Reddit overall, clearly state that consent must be obtained for any content posted \cite{reddit-rules}. Yet, our manual analysis identified multiple forms of non-consensual content sharing: 

\circled{1} \textbf{Requests for identification}: Posts that explicitly ask for help in identifying someone in a photograph or video (``Does anyone know who this is?'' or ``Help to identify this wife'').
\circled{2} \textbf{Mentions of unawareness}: Posts that mention the subject being unaware of the content being shared (``She doesn’t know I’m posting this,'' or ``got access to her phone'') are strong indicators of non-consensual sharing, highlighting the lack of consent and awareness of the subject, making the content potentially harmful and unethical.
\circled{3} \textbf{Mentions of leaked content}:  Mentions of ``leaked content'' often signal the illicit release of private material, typically without the subject's consent.
\circled{4} \textbf{Mentions of hacking}: References to ``hacking'' in online discourse frequently denote activities that compromise personal accounts or digital privacy, leading to unauthorized access and distribution of content.

\circled{5} \textbf{Mentions of under-age content}: References to explicit content involving individuals under the age of 18 are considered illegal and constitute a violation of platform policies; such content is therefore classified as non-consensual.

To generalize beyond our manual analysis, we trained a classifier that is able to detect such posts from our examples. 


\subsection{Classifier training}
\label{sec:classifier_new}
We conduct experiments using five widely utilized machine learning models for text classification tasks to identify posts indicative of non-consensual sharing. We utilize our labeled data (298 posts labeled non-consensual; 3420 as consensual) for building the classifiers. The task is formulated as a binary classification problem, distinguishing between non-consensual and consensual sharing. The models are trained on 70\% of the data, evaluated on 15\% data, and the remaining 15\% is held out for final assessment. We use NVIDIA Titan RTX GPU with 24 GB of memory for training. We experiment with the following models:

\begin{enumerate}
    \item Random Forest with TF-IDF representation, implemented using the scikit-learn package, and parameters optimized using GridSearchCV with 5-fold cross-validation.
    
    \item Logistic Regression, also built using scikit-learn, leveraging GridSearchCV with 5-fold cross-validation for parameter optimization.

    \item  GPT-4 with a zero-shot learning approach for classification through instruction prompting. We design a prompt that instructs the model on the classification objective and defines the desired output format (full details given in Appendix \ref{app:prompt})

    \item A fine-tuned RoBERTa-Large model with cross entropy loss (Roberta-cross)  over 5 training epochs with a batch size of 16, a learning rate of $1e^{-5}$ and the AdamW optimizer.
    
    \item A fine-tuned RoBERTa-Large model with supervised contrastive learning (Roberta-scl) \cite{gunel2020supervised} with the same parameters.
    
\end{enumerate}

\subsection{Results}
\label{sec:classifier_results}
Table~\ref{tab:metric} presents the results.
Roberta-based classifiers outperform other classifiers in all evaluation metrics except GPT-4 for Recall. The higher Recall exhibited by GPT-4 indicates that it is more likely to classify negative instances as positive, leading to a greater number of false positives reflected by the lowest value of precision. While a high recall rate can be valuable in ensuring that the most relevant instances are detected, it comes with a trade-off in the context of NSFW subreddits. Specifically, this tendency towards false positives means that GPT-4 may incorrectly flag or block posts that actually comply with the subreddit's guidelines. This could result in unintended censorship of permissible content, undermining the intended purpose of the community and potentially excluding posts that align with the subreddit's rules. However, this could still be useful with manual oversight, as we discuss below.

\begin{figure}[ht]
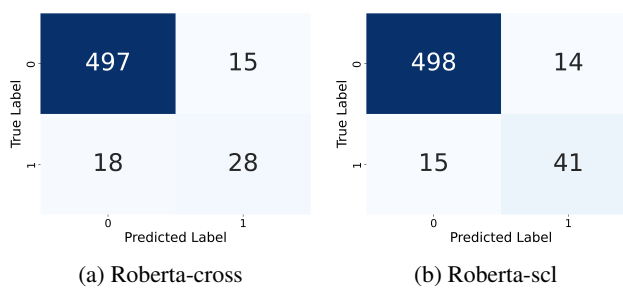

    \centering
    \begin{subfigure}[b]{0.23\textwidth}
        \centering
        \includegraphics[width=\textwidth]{Figures/CRV_Figures/confusion_cross_updated.pdf}
        \caption{Roberta-cross}
        \label{fig:conf_cross}
    \end{subfigure}
    \hfill
    \begin{subfigure}[b]{0.23\textwidth}
        \centering
        \includegraphics[width=\textwidth]{Figures/CRV_Figures/confusion_scl_updated.pdf}
        \caption{Roberta-scl}
        \label{fig:conf_scl}
    \end{subfigure}
    \caption{Confusion matrix on manually checked 500 posts}
    \label{fig:conf_mat}
\end{figure}

 Roberta-cross achieves a higher recall compared to Roberta-scl, whereas Roberta-scl outperforms Roberta-cross in terms of precision. Given the limited size of the holdout dataset, drawing definitive conclusions about which model performs better is challenging. Therefore, to evaluate the models' performance on new data, we sampled 5\% of the posts from each remaining 10 subreddits, resulting in a test set of  217,968 posts. We then applied the models model to this dataset. We manually evaluate 500 random posts from this data, to see whether the models correctly classify them. We find that Roberta-scl (Recall=0.89) outperforms the Roberta-cross classifier (Recall=0.60). Supervised contrastive learning pushes the embeddings of data points belonging to the same class close and of different classes apart. Figure~\ref{fig:conf_mat} presents the confusion matrix of both classifiers on 500 manually checked posts.

\noindent\textbf{Suggested Directions}: As NSFW content starts to scale in volume with the new policies on platforms such as X, we believe that classifiers such as the one we have developed will be an essential component in the toolkit of human moderators. We present our initial models only as a baseline for what is possible and advocate further research into identifying better tools and methods to assist moderators in maintaining healthy online conversations.

Given the importance of detecting non-consensual practices online, we believe that models with higher recall are more valuable, to ensure the model minimizes missed positive instances.   Such a classifier can then be used as a tool that can \textit{assist} moderators of NSFW platforms. With human oversight, the moderators can be in control of deciding whether or not something is \textit{really} a false positive, ensuring accurate identification of potential violations. It is important to retain human oversight as completely automated methods could inadvertently censor wrong posts, or miss-identify truly non-consensual ones.

\begin{table}[]
\caption{Performance of classifiers on held out dataset}
\label{tab:metric}
\begin{tabular}{@{}lrrrr@{}}
\toprule
\textbf{Metrics}                                                        & \textbf{Accuracy} & \textbf{Precision} & \textbf{Recall} & \textbf{F1-score} \\ \midrule
\textbf{\begin{tabular}[c]{@{}l@{}}Random\\ Forest\end{tabular}}        & 0.96              & 0.66               & 0.59            & 0.61              \\
\textbf{\begin{tabular}[c]{@{}l@{}}Logistic \\ Regression\end{tabular}} & 0.95              & 0.67               & 0.48            & 0.56              \\
\textbf{Gpt-4}                                                          & 0.75              & \textbf{0.52}      & 0.95            & 0.67              \\
\textbf{\begin{tabular}[c]{@{}l@{}}RoBERTa-\\ cross\end{tabular}}       & 0.98              & 0.76               & \textbf{0.98}   & 0.86              \\ 
\textbf{\begin{tabular}[c]{@{}l@{}}RoBERTa-\\ scl\end{tabular}}         & 0.98              & \textbf{0.72}      & 0.86            & 0.77              \\ \bottomrule
\end{tabular}
\end{table}

\section{Conclusion}
\label{sec:conclusion}

Not Safe For Work (NSFW) content has long been a sensitive category around which there are prohibitions or restrictions on many mainstream social platforms. However, recently there has been a shift in this position, with X relaxing its rules and permitting consensual sharing of NSFW content~\cite{Xpolicy}. To understand how this might play out, we turn to Reddit, one of the few mainstream platforms that has allowed sharing of NSFW content on clearly marked subreddits. In particular, we focus on NSFW subreddits which are \textit{restricted}, allowing only vetted users to share content, which should lead to responsible sharing of NSFW content, including respecting rules around consent.

Interestingly, we found that some of the approved posters on such restricted subreddits are using it as a springboard for drawing users to other platforms such as OnlyFans or Kik, where such content can be monetized. Others are using payment mechanisms such as Venmo, Apple Cash, or cryptocurrency to directly trade on the platform, obtaining up to \$1300. Many such users skirting the rules use throwaway accounts, which enable them to rejoin with a different user ID if banned. Throwaway accounts are also used for coordinated advertisement campaigns, violating the sharing spirit of the community.

As our exploration progressed, we also identified signs suggesting that not all content shared appeared to be fully consensual. Certain posts included language or comments that hinted at one partner’s unawareness, sparking ethical questions about the boundaries of consensual sharing in online spaces. This revelation prompted a deeper investigation into the nature of posts to understand the extent of non-consensual content that may be circulating, the language patterns associated with these practices, and the implications for both users and platform regulators. To this end, we developed a RoBERTa-based model to identify non-consensual posts. As mandated by our institution's ethics committee, we have reported these posts to a relevant charity as well as law enforcement authorities. We have also reached out to the moderators of the corresponding subreddits, and have provided them access to our model to assist them in enforcing consent-related guidelines.

\noindent\textbf{Discussion}: Through this focused study of the largest (restricted) NSFW subreddits, each with a subscriber base of over 1 million users, we aim to contribute to the broader discourse on ethical content sharing and moderation within online communities. Although NSFW subreddits typically intend to \textit{share} NSFW content freely and consensually, commerce and monetization become an important angle, changing the character of the community. Consent, which is an important ground rule on all of Reddit~\cite{reddit-rules},  appears to have been ignored in a small but significant number of cases. This has obvious implications for other mainstream platforms such as X, which are now starting to allow NSFW content. If NSFW content enters mainstream, it may present new monetization opportunities for platforms, but the impact on the rest of the platform and its character should be considered. It will become even more important to ensure consent, since it may be easier to detect NSFW content than to determine whether such content was always shared in a legal and consensual manner.

\bibliography{nsfw}

\appendix

\section{APPENDIX}
\subsection{Prompts for GPT-4}
\label{app:prompt}
\subsubsection{Prompt for username extraction:}

few\_shot\_template:
\\
\vspace{2pt}
Post: Anyone want to trade their hot Asian gf for my hot Asian ex gf or current gf ? r*ib*i3*0* add up
\\
\vspace{2pt}
Q: What is the username mentioned in the post?
\\
\vspace{2pt}
A: r*ib*i3*0*
\\
\vspace{2pt}
Post: Only 9 more discounted subs left! 50\% off for full access to SOLO-MF-FFM-FF-FFFFM content LINKS ON MY PROFILE AND FREE EXCLUSIVE TEASERS ON MY SNAPCHAT-Pe*c*es*D*G
\\
\vspace{2pt}
Q: What is the username mentioned in the post?
\\
\vspace{2pt}
A: Pe*c*es*D*G
\\
\vspace{2pt}

Post: {post}
\\
\vspace{2pt}
Q: What is the username mentioned in the post?
\vspace{2pt}
A:"""
   
\subsubsection{Prompt for classification task:}
Instruction prompt: Classify the following text as 0 or 1:
Text: "{text}"
Your task is to determine whether the text is indicative of non-consensual image sharing.
Non-consensual image sharing refers to the distribution or sharing of images, videos, or other media without the consent of the person depicted. If the text is indicative of non-consensual image sharing, output 1. Otherwise, output 0.  Your answer should be either 0 or 1

\section{Ethics Statement}
Analyzing NSFW content and detecting non-consensual sharing is a highly sensitive subject. This task presents several challenges, including safeguarding user privacy, ensuring the comfort and ethical considerations of researchers, and mitigating potential risks associated with handling such data. To ensure the fairness and reliability of our work, we secured approval from the Institutional Ethics Board (IRB) for this research. Our ethics statement outlines the following essential ethical considerations:
\begin{itemize}
    \item \textbf{Confidentiality:} To safeguard the privacy of Reddit users, we refrained from performing any cross-platform linking. Additionally, our analysis did not focus on user-specific patterns but instead centered on a general examination of NSFW subreddit. We did not collect any videos or images associated with the posts. To ensure the privacy of Reddit users, our study was strictly limited to the text of posts and associated metadata.
    \item \textbf{Potential for harm:} We implement a protocol to notify competent authorities to take action in case we observe any illegal content within the subreddit. Our aim is not to harm Reddit users in any way. Our ethics statement ensures that our reporting focuses solely on identifying potential posts suggesting non-consensual sharing, without disclosing any Reddit user IDs or post IDs.
    \item \textbf{Results communication:} We remain mindful in sharing our research findings ensuring that our research is presented responsibly, avoiding unnecessary sensationalism and does not include any identifying information. We are not reporting any personally identifiable information (PII) in our report to ensure that the data cannot be linked back to individual users.
\end{itemize}

\section{Paper Checklist}
\subsection{Paper Checklist to be included in your paper}

\begin{enumerate}
\item For most authors...
\begin{enumerate}
    \item  Would answering this research question advance science without violating social contracts, such as violating privacy norms, perpetuating unfair profiling, exacerbating the socio-economic divide, or implying disrespect to societies or cultures?
    \answerTODO{Yes, we have discussed this in our Ethics Statement.}
  \item Do your main claims in the abstract and introduction accurately reflect the paper's contributions and scope?
    \answerTODO{Yes, we have discussed the main contributions and
findings of the paper in the abstract and introduction.}
   \item Do you clarify how the proposed methodological approach is appropriate for the claims made? 
    \answerTODO{Yes, we have described it in section \ref{sec:data}}
   \item Do you clarify what are possible artifacts in the data used, given population-specific distributions?
    \answerTODO{Yes, section \ref{sec:data} presents the descriptive analysis of data.}
  \item Did you describe the limitations of your work?
    \answerTODO{Yes, we described it in section \ref{sec:conclusion}}
  \item Did you discuss any potential negative societal impacts of your work?
    \answerTODO{Yes, we have discussed this in
Ethical Statement section.}
      \item Did you discuss any potential misuse of your work?
    \answerTODO{Yes, we have discussed this in
Ethical Statement section.}
    \item Did you describe steps taken to prevent or mitigate potential negative outcomes of the research, such as data and model documentation, data anonymization, responsible release, access control, and the reproducibility of findings?
    \answerTODO{Yes, we have discussed this in
Ethical Statement and also.}
  \item Have you read the ethics review guidelines and ensured that your paper conforms to them?
    \answerTODO{Yes}
\end{enumerate}

\item Additionally, if your study involves hypotheses testing...
\begin{enumerate}
  \item Did you clearly state the assumptions underlying all theoretical results?
    \answerTODO{NA}
  \item Have you provided justifications for all theoretical results?
    \answerTODO{NA}
  \item Did you discuss competing hypotheses or theories that might challenge or complement your theoretical results?
    \answerTODO{NA}
  \item Have you considered alternative mechanisms or explanations that might account for the same outcomes observed in your study?
    \answerTODO{NA}
  \item Did you address potential biases or limitations in your theoretical framework?
    \answerTODO{NA}
  \item Have you related your theoretical results to the existing literature in social science?
    \answerTODO{NA}
  \item Did you discuss the implications of your theoretical results for policy, practice, or further research in the social science domain?
    \answerTODO{NA}
\end{enumerate}

\item Additionally, if you are including theoretical proofs...
\begin{enumerate}
  \item Did you state the full set of assumptions of all theoretical results?
    \answerTODO{NA}
	\item Did you include complete proofs of all theoretical results?
    \answerTODO{NA}
\end{enumerate}

\item Additionally, if you ran machine learning experiments...
\begin{enumerate}
  \item Did you include the code, data, and instructions needed to reproduce the main experimental results (either in the supplemental material or as a URL)?
    \answerTODO{Yes,we have added all the relevant details for reproducibility
and further research. We will make model code publicly available upon the paper’s acceptance.}
  \item Did you specify all the training details (e.g., data splits, hyperparameters, how they were chosen)?
    \answerTODO{Yes, they are given in section \ref{sec:classifier_new} }
     \item Did you report error bars (e.g., with respect to the random seed after running experiments multiple times)?
    \answerTODO{No, we have fixed the seed value to ensure the reproducibility of research.}
	\item Did you include the total amount of compute and the type of resources used (e.g., type of GPUs, internal cluster, or cloud provider)?
    \answerTODO{Yes, we mentioned it in section \ref{sec:classifier_new}}
     \item Do you justify how the proposed evaluation is sufficient and appropriate to the claims made? 
    \answerTODO{Yes, we have described it in section \ref{sec:classifier_new}}
     \item Do you discuss what is ``the cost`` of misclassification and fault (in)tolerance?
    \answerTODO{Yes, in section section \ref{sec:classifier_new}}
  
\end{enumerate}

\item Additionally, if you are using existing assets (e.g., code, data, models) or curating/releasing new assets, \textbf{without compromising anonymity}...
\begin{enumerate}
  \item If your work uses existing assets, did you cite the creators?
    \answerTODO{Yes, we have cited relevant papers.}
  \item Did you mention the license of the assets?
    \answerTODO{NA}
  \item Did you include any new assets in the supplemental material or as a URL?
    \answerTODO{NA}
  \item Did you discuss whether and how consent was obtained from people whose data you're using/curating?
    \answerTODO{NA }
  \item Did you discuss whether the data you are using/curating contains personally identifiable information or offensive content?
    \answerTODO{Yes, we discuss this in our ethics statement}
\item If you are curating or releasing new datasets, did you discuss how you intend to make your datasets FAIR?
\answerTODO{Our ethics protocol does not allow the release of dataset considering the sensitivity of dataset.}
\item If you are curating or releasing new datasets, did you create a Datasheet for the Dataset (see \citet{gebru2021datasheets})? 
\answerTODO{NA}
\end{enumerate}

\item Additionally, if you used crowdsourcing or conducted research with human subjects, \textbf{without compromising anonymity}...
\begin{enumerate}
  \item Did you include the full text of instructions given to participants and screenshots?
    \answerTODO{NA}
  \item Did you describe any potential participant risks, with mentions of Institutional Review Board (IRB) approvals?
    \answerTODO{NA}
  \item Did you include the estimated hourly wage paid to participants and the total amount spent on participant compensation?
    \answerTODO{NA}
   \item Did you discuss how data is stored, shared, and deidentified?
   \answerTODO{NA}
\end{enumerate}

\end{enumerate}

\end{document}